\documentclass[twocolumn,showpacs,aps,prl,superscriptaddress]{revtex4}

\usepackage[dvips]{graphicx}
\usepackage{latexsym}

\begin{document}

\title{Virtual Compton Scattering in the Resonance Region up to the Deep
       Inelastic Region at Backward Angles and Momentum Transfer Squared
       of $Q^2=1.0$~GeV$^2$}

\author{G.~Laveissi\`{e}re}
\affiliation{Universit\'{e} Blaise Pascal/IN2P3, F-63177 Aubi\`{e}re, France}
\author{N.~Degrande}
\affiliation{University of Gent, B-9000 Gent, Belgium}
\author{S.~Jaminion}
\affiliation{Universit\'{e} Blaise Pascal/IN2P3, F-63177 Aubi\`{e}re, France}
\author{C.~Jutier}
\affiliation{Universit\'{e} Blaise Pascal/IN2P3, F-63177 Aubi\`{e}re, France}
\affiliation{Old Dominion University, Norfolk, VA 23529}
\author{L.~Todor}
\affiliation{Old Dominion University, Norfolk, VA 23529}
\author{R.~Di Salvo}
\affiliation{Universit\'{e} Blaise Pascal/IN2P3, F-63177 Aubi\`{e}re, France}
\author{L.~Van Hoorebeke}
\affiliation{University of Gent, B-9000 Gent, Belgium}
\author{L.C.~Alexa}
\affiliation{University of Regina, Regina, SK S4S OA2, Canada}
\author{B.D.~Anderson}
\affiliation{Kent State University, Kent OH 44242}
\author{K.A.~Aniol}
\affiliation{California State University, Los Angeles, Los Angeles, CA 90032}
% ...
\collaboration{The Jefferson Lab Hall A Collaboration}
\noaffiliation

\makeatletter
\global\@specialpagefalse
\def\@oddhead{\hfill {G. Laveissiere {\it et al.}} {photon electroproduction}}
\let\@evenhead\@oddhead
\def\@oddfoot{\reset@font\rm\hfill \thepage\hfill
} \let\@evenfoot\@oddfoot
\makeatother

\begin{abstract}
We have made the first measurements of the virtual Compton scattering process
via the $e p \rightarrow e p \gamma$ exclusive reaction at $Q^2 = 1$~GeV$^2$
in the nucleon resonance region.
The cross section is obtained at center of mass (CM) backward angle,
in a range of total $(\gamma^* p)$ CM energy $W$ from the proton mass up to
$W = 1.91$~GeV. The data show resonant structures in the first and second
resonance regions, and are well reproduced at higher $W$ by the Bethe-Heitler+Born
cross section, including t-channel $\pi^0$-exchange.
At high $W$, our data, together with existing real photon data, show a striking
$Q^2$ independence.
Our measurement of the ratio of $H(e,e'p)\gamma$ to $H(e,e'p)\pi^0$ cross sections
is presented and compared to model predictions.
\end{abstract}


\maketitle

Understanding nucleon structure in terms of the non-perturbative dynamics
of quarks and gluons requires new and diverse experimental data to guide
theoretical approaches and to constrain models. Purely electro-weak processes
are privileged tools since they can be interpreted directly in terms of the
current carried by the quarks.
This Letter presents a study of the virtual Compton scattering (VCS) process~:
$\gamma^\star p \rightarrow \gamma p$, in the nucleon resonance region via the
photon electroproduction reaction~: $H(e,e'p)\gamma$. For the first time,
we separate this process from the dominant
$H(e,e'p)\pi^0$ reaction above pion threshold.

The Constituent Quark Model of Isgur and Karl~\cite{Isgur:1979wd}
reproduces many features of the nucleon spectrum. However, for $W>1.6$~GeV,
the model predicts a number of positive parity resonances~\cite{Koniuk:1980vw}
that have not been seen experimentally.
The simultaneous study of both $(N\pi)$ and $(N\gamma)$ final states of the
electroproduction process on the nucleon offers probes with very different
sensitivities to the resonance structures.
Another motivation for the present study is to explore the exclusive
$e p \rightarrow e p \gamma$ reaction at high $W$, where current quark
degrees of freedom may become as important as those of constituent quarks
in the understanding of resonances. Quark-hadron duality implies that even at modest
$Q^2$, inelastic electron scattering in the resonance region can be analyzed
in terms of quark degrees of freedom in the $t$-channel of the forward
Compton amplitude instead of nucleon resonances in the $s$-channel~\cite{Niculescu:2000tk}.

\begin{figure}[ht]
\includegraphics[width=\linewidth]{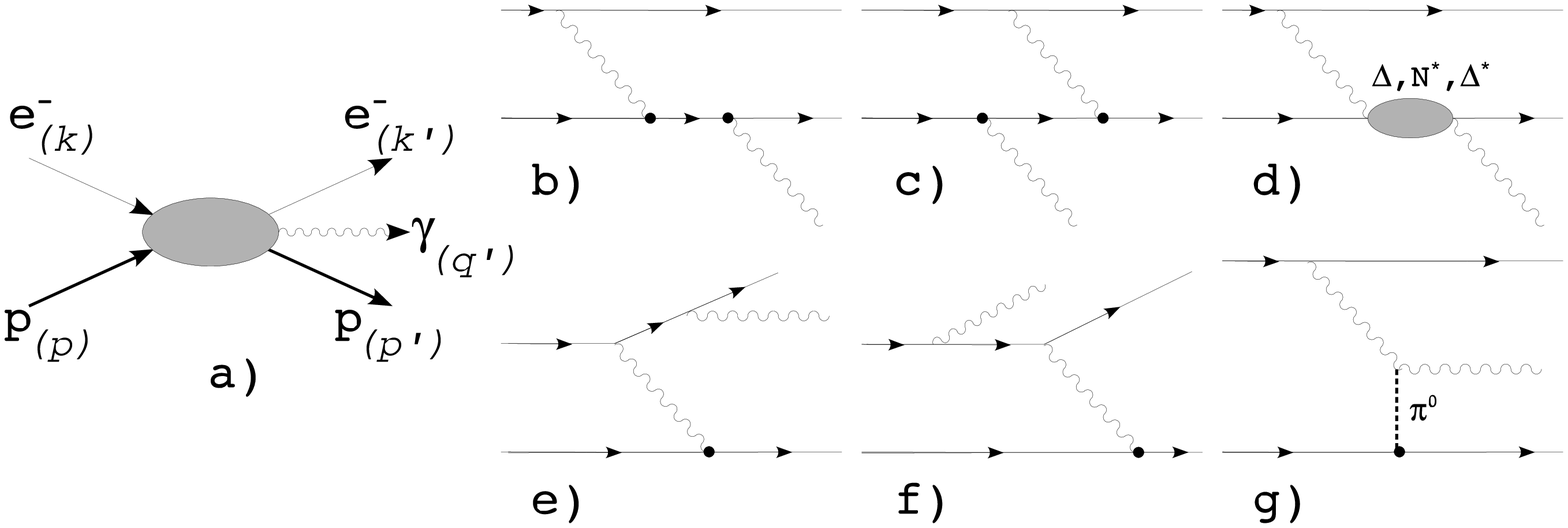}
\caption{\label{fig:feynman}Kinematics for photon electroproduction on the
         proton (a) and lowest order amplitudes for VCS Born ($\rm{b}$, $\rm{c}$),
         VCS Non-Born (d), Bethe-Heitler ($\rm{e}$, $\rm{f}$) and $t$-channel
         $\pi^0$-exchange (g)~\cite{Vanderhaeghen:1996iz} processes. The particles
         4-momenta are indicated in parenthesis.}
\end{figure}
We define the kinematics of the $e p \rightarrow e p \gamma$ reaction in
Fig.~\ref{fig:feynman}a. A common set of invariant kinematic variables is
defined as $-Q^2=(k-k')^2=q^2$, $s=W^2 = (q+p)^2$, $t=(p-p')^2$,
and $u=(p-q')^2$.
The $\vec{q}$-direction defines the polar-axis of the coordinate system~:
$\theta_{\gamma\gamma}^\ast$ and $\phi_{\gamma\gamma}$
are the polar and azimuthal angles in the $\gamma^\star p \rightarrow \gamma p$
subprocess CM frame. The scattered electron direction defines
$\phi_{\gamma\gamma}=0$.
The $e p \rightarrow e p \gamma$ reaction was measured below pion threshold
at MAMI ($Q^2=0.33$~GeV$^2$)~\cite{Roche:2000ng} and at the Thomas Jefferson
National Accelerator Facility (JLab) ($Q^2=0.92$ and $1.76$~GeV$^2$)~\cite{Laveissiere:2004nf}.
We present in this letter the first measurements of the nucleon excitation up
to $W=1.91$~GeV at $Q^2=1$~GeV$^2$ through the $e p \rightarrow e p \gamma$
process in CM backward kinematics ($\vec{q}^{\,\prime}$ opposite to $\vec{q}$).

In the one photon exchange approximation, the $e p \rightarrow e p \gamma$
amplitude (Fig.~\ref{fig:feynman}a) includes the coherent superposition of
the VCS Born (Fig.~\ref{fig:feynman}$\rm{b}$
and \ref{fig:feynman}$\rm{c}$) and Non-Born (Fig.~\ref{fig:feynman}d) amplitudes,
and the Bethe-Heitler (BH) amplitude (Fig.~\ref{fig:feynman}$\rm{e}$ and
\ref{fig:feynman}$\rm{f}$)~\cite{Bethe:1934za}.
Note that in the BH amplitude, the mass of the virtual photon (elastically
absorbed by the proton) is $t$. In the VCS amplitude, the mass of the
virtual photon (inelastically absorbed) is $-Q^2$.
The BH amplitude dominates over the VCS when the photon is emitted in either
the direction of the incident or scattered electron, and breaks the symmetry of the
electroproduction amplitude around the virtual photon direction.
Thus, it is not possible to expand the $\phi_{\gamma\gamma}$-dependence of the
$e p \rightarrow e p \gamma$ cross section in terms of the usual electroproduction
$LT$ and $TT$ interference terms, except when the BH amplitude is really negligible.
Above the $\Delta$-resonance, the VCS amplitude is the dominant contribution in our
kinematics.
In this region, the data are $\phi_{\gamma\gamma}$-independent, within our statistics.
For these reasons, we have not performed azimuthal analysis of the data.

\begin{figure}[t]
\includegraphics[width=\linewidth]{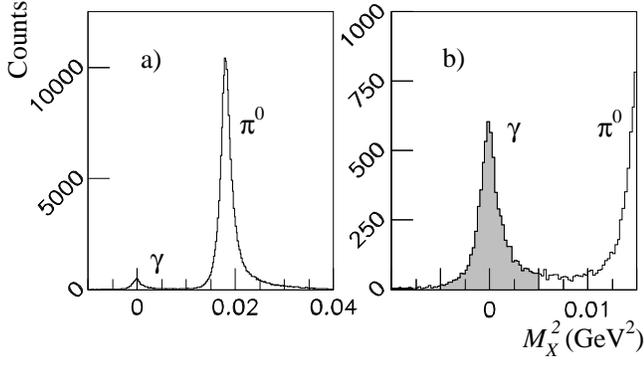}
\caption{\label{fig:missing-mass}Squared missing mass $M_X^2$
         for an experimental setting at $W=1.2$~GeV (a) and zoom around the
         $\gamma$ peak (b). The shaded window [--0.005,0.005~GeV$^2$] is used to
         select the $\gamma$ events.}
\end{figure}

The experiment was performed at JLab in Hall~A. The continuous electron beam of
energy $4.032$~GeV with an intensity of 60-120~$\mu$A
bombarded a 15~cm liquid hydrogen target.
The scattered electron and
recoil proton were detected in coincidence in two high resolution spectrometers.
The emitted photon
was reconstructed using a missing particle technique. A spectrum of the squared
missing mass $M_X^2=(k+p-k'-p')^2$ is displayed in Fig~\ref{fig:missing-mass}.
We give a detailed description of the apparatus, detector acceptance, and systematic
errors in~\cite{Alcorn:2004}, \cite{Laveissiere:2003jf} and \cite{Laveissiere:2001th}.

\begin{figure}[ht]
\includegraphics[width=\linewidth]{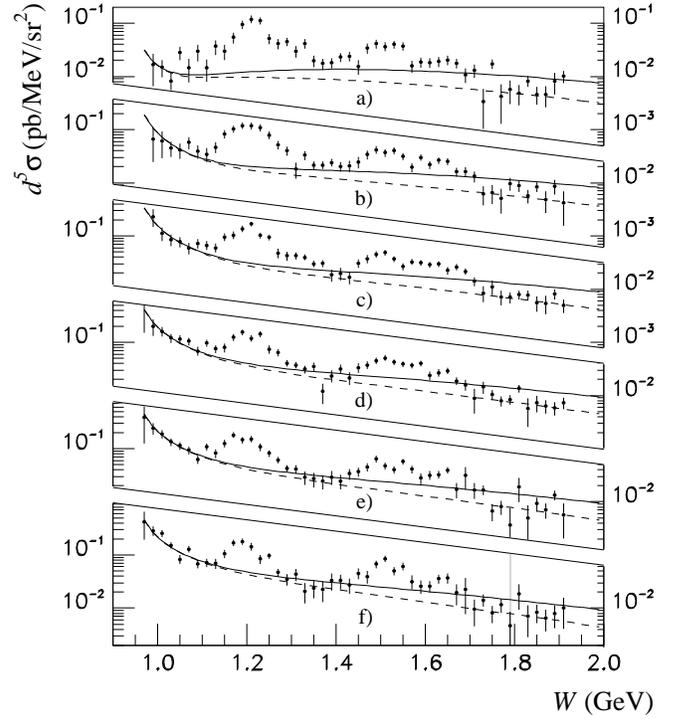}
\caption{\label{fig:excitation-curve}Excitation curves for $e p \rightarrow
         e p \gamma$ at $Q^2=1$~GeV$^2$, $\cos\theta_{\gamma\gamma}^\ast =-0.975$
         and $\phi_{\gamma\gamma}=15^\circ$ (a), $45^\circ$ (b), $75^\circ$ (c),
         $105^\circ$ (d), $135^\circ$ (e) and $165^\circ$ (f). The full line is the
         BH+Born cross section, and the dashed line is the BH+Born+$\pi^0$-exchange
         cross section~\cite{Vanderhaeghen:1996iz} (see text).}
\end{figure}

\begin{figure}[t]
\includegraphics[width=\linewidth]{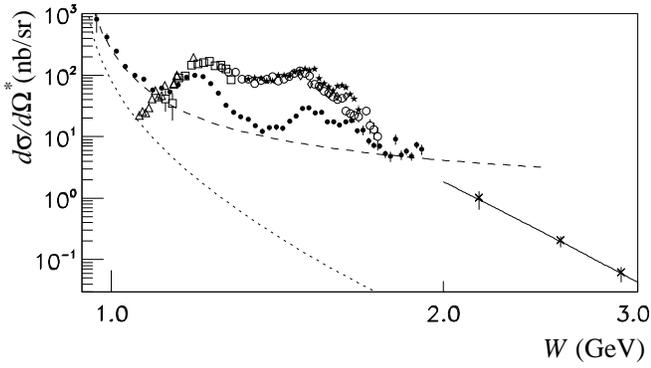}
\caption{\label{fig:rcs-vcs}Comparison of VCS data from this experiment ($\bullet$)
         at $\theta_{\gamma\gamma}^\ast = 167.2^\circ$ and RCS data at
         $\theta_{\gamma\gamma}^\ast = 159-162^\circ$ ($\star$)~\cite{Wada:1984sh},
         $\theta_{\gamma\gamma}^\ast = 128-132^\circ$ ($\diamondsuit$)~\cite{Jung:1981wm},
         $\theta_{\gamma\gamma}^\ast = 141^\circ$     ($\triangle$)~\cite{Hallin:1993ft},
         $\theta_{\gamma\gamma}^\ast = 130-133^\circ$ ($\circ$)~\cite{Ishii:1985ei},
         $\theta_{\gamma\gamma}^\ast = 131^\circ$     ($\Box$)~\cite{Wissmann:1999vi} and
         $\theta_{\gamma\gamma}^\ast = 105-128^\circ$ ($\times$)~\cite{Shupe:1979vg}).
         The solid curve is an $s^{-6}$ power law normalized to the
         $W=2.55$~GeV Cornell point.
         The dashed curve is the BH+Born+$\pi^0$-exchange cross section (see text)
         and the dotted curve the BH alone.}
\end{figure}

\begin{figure}[t]
\includegraphics[width=\linewidth]{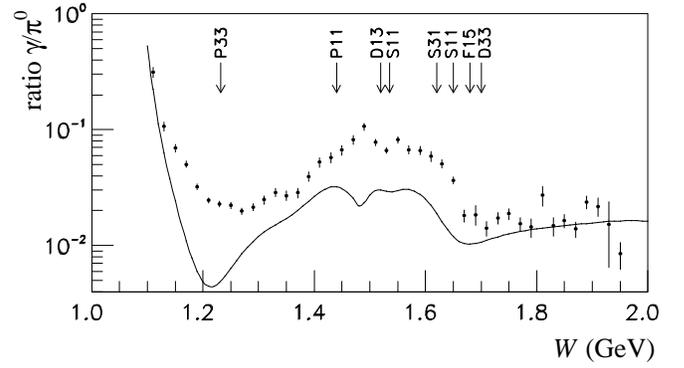}
\caption{\label{fig:ratio-gam-pi0}Ratio of $e p \rightarrow e p \gamma$ over
         $e p \rightarrow e p \pi^0$ cross sections at $Q^2=1$~GeV$^2$ and
         $\theta_{\gamma\gamma}^\ast = 167.2^\circ$.
         The full curve is the ratio between the BH+Born+$\pi^0$-exchange
         cross section and the MAID2003 model~\cite{Drechsel:1998hk}.}
\end{figure}

We extract the 5-fold differential cross section
$d^5\sigma=d^5\sigma/dk'_{lab}d[\Omega_e]_{lab}d[\Omega_p]_{CM}$
using the method described in~\cite{Laveissiere:2003jf};
$dk'_{lab}$ and $d[\Omega_e]_{lab}$ are the scattered electron differential
momentum and solid angle in the lab frame, and $d[\Omega_p]_{CM}$ is the proton CM
differential solid angle.
The calculations of the solid angle and radiative corrections
are based on a simulation~\cite{VanHoorebeke} including the coherent sum
of the BH and VCS-Born amplitudes (Fig.~\ref{fig:feynman}$\rm{b}$, $\rm{c}$,
$\rm{e}$ and $\rm{f}$) only.
The inclusion of the BH-amplitude ensures that our simulation reproduces the
strong $\phi_{\gamma\gamma}$ dependence near threshold.
Corrections were applied for acceptance, trigger efficiency, acquisition and
electronic dead times, tracking efficiency, target boiling, target impurity
and proton absorption~\cite{Laveissiere:2003jf}.
In addition, the correction ($-0.1$ to $-1.7\%$) for exclusive $\pi^0$
background in the $M_X^2$ window $[-0.005, 0.005]$~GeV$^2$ was made using
our simulation~\cite{Laveissiere:2003jf}.
%
The main results are displayed in Fig.~\ref{fig:excitation-curve}, \ref{fig:rcs-vcs}
and \ref{fig:ratio-gam-pi0}. On these figures, errors are statistical only; the
systematic errors are of the same order.

In Fig.~\ref{fig:excitation-curve} we present the 5-fold differential cross
sections $d^5\sigma$ for the six bins in $\phi_{\gamma\gamma}$ ($30^\circ$
wide) as a function of $W$, at $Q^2=1$~GeV$^2$ and $\cos\theta_{\gamma\gamma}^\ast = -0.975$.
By symmetry, the statistics from $\phi_{\gamma\gamma}=0$ to $-180^\circ$ are also included.
The cross sections we obtain are displayed in tables~\ref{table:ctcm1}, \ref{table:ctcm2}
and \ref{table:ctcm3}.
Note that these tables are not completely filled due to the acceptance of the experiment.
The data show strong resonance phenomena in the first and second resonance regions,
but the higher resonances are not distinguishable due to the limited statistical precision.

The VCS data in Fig:~\ref{fig:excitation-curve} approach the BH+Born
calculation at high $W$, when we include the destructive $t$-channel $\pi^0$ exchange
graph~\cite{Vanderhaeghen:1996iz} of Fig.~\ref{fig:feynman}g).
In the resonance model of Capstick and Keister~\cite{Capstick:1993qi},
the positive parity intermediate states contribute constructively and the negative
parity states contribute destructively to the backward angle cross section.
If there are
no diffractive minima as a function of $Q^2$, this effect will remain in VCS.
As the level density grows with $W$, the positive and negative parity resonances tend to
compensate and the backward
cross section decreases with increasing $W$.

Real Compton scattering (RCS) has been intensively investigated in the
$\Delta(1232)$-resonance~\cite{Wissmann:1999vi} and in the high energy diffractive
region~\cite{Bauer:1978iq}. RCS was also studied above
the $\Delta(1232)$ at Bonn~\cite{Jung:1981wm}, Saskatoon~\cite{Hallin:1993ft}, and
Tokyo~\cite{Wada:1984sh,Ishii:1985ei}. The Cornell experiment~\cite{Shupe:1979vg}
measured the RCS process at photon energies $E_\gamma$ in the range 2--6~GeV and angles
from $45^\circ$ to $128^\circ$ in the CM frame. The recent JLab
experiment E99-114 measured the RCS process at $E_\gamma$ in the range 3--6~GeV.
We note that the large angle RCS data are roughly independent of angle for 
$\theta_{\gamma\gamma}^\ast > 90^\circ$. The Cornell RCS data ($W$, $-t$, $-u$ all large) are 
qualitatively
consistent with the $s^{-6}$ scaling law for $d\sigma/dt$~\cite{Brodsky:1975vy}
or with a $t$-dependent Compton form factor~\cite{Radyushkin:1998rt,Huang:2001ej,Miller:2004rc}.
In Fig.~\ref{fig:rcs-vcs}, we compare our backward angle $e p \rightarrow e p \gamma$
cross section divided by the photon flux factor (Hand convention~\cite{Laveissiere:2003jf})
with existing large angle RCS data. For comparison purposes, we average our VCS cross sections 
over the azimuthal angle $\phi_{\gamma\gamma}$.

In the resonance region, the VCS/RCS comparison in Fig.~\ref{fig:rcs-vcs} shows a strong decrease
with $Q^2$ as expected from resonance form factors.
This is in contrast with the behavior in the region of
$W\sim 1.8$~GeV where our VCS data are approximately equal to the wide angle RCS data.
This evidence of $Q^2$-independence suggests that the Compton process is coupling
to point-like constituents.
This conjecture can be tested with new backward angle VCS data at large $W$ 
and at this $Q^2$ and above.  If this $Q^2$-independence is confirmed, we emphasize that this is a new 
probe of the quark structure of the proton.

From these results and those presented in~\cite{Laveissiere:2003jf}, we have
computed the ratio between the $e p \rightarrow e p \gamma$ and $e p \rightarrow e p \pi^0$
cross sections at $\cos\theta_{\gamma\gamma}^\ast = -0.975$ and
$Q^2=1$~GeV$^2$ for the entire resonance region.
In Fig.~\ref{fig:ratio-gam-pi0}
we show the value of the ratio averaged over the six bins in $\phi_{\gamma\gamma}$.
In the region of the $P_{33}$ $\Delta$-resonance, the experimental ratio is five times
larger than the BH+Born+$\pi^0$-exchange cross section divided by the
MAID2003~\cite{Drechsel:1998hk} calculation.
Thus one has to be careful when correcting $H(e,e'p)\pi^0$ experiments if the resolution
does not allow a clear separation of the $H(e,e'p)\gamma$ channel.

In summary, we studied for the first time the process $e p \rightarrow e p \gamma$
in the nucleon resonance region. The data is dominated by the VCS process and show
similar resonance phenomena as for RCS.
At high $W$ the $Q^2$ independence of the data suggest the VCS process is coupling
to elementary quarks. The comparison with $e p\rightarrow e p \pi^0$ shows strong
variations with $W$ across the resonance region.

\begin{acknowledgments}
We wish to acknowledge essential work of the JLab accelerator staff
and Hall A technical staff.
This work was supported by DOE contract DE-AC05-84ER40150 under
which the Southeastern Universities Research Association (SURA)
operates the Thomas Jefferson National Accelerator Facility. We
acknowledge additional grants from the US DOE and NSF, the French
Centre National de la Recherche Scientifique and Commissariat \`a
l'Energie Atomique, the Conseil R\'egional d'Auvergne, the
FWO-Flanders (Belgium) and the BOF-Gent University.
\end{acknowledgments}

\begingroup\squeezetable
\begin{table*}\begin{ruledtabular}
\caption{\label{table:ctcm1}$e p \rightarrow e p \gamma$ cross section ($\pm$ stat $\pm$ syst)
         at $Q^2$=1.0~GeV$^2$ and $\cos\theta_{\gamma\gamma}^\ast = -0.975$, in fm/MeV/sr$^2$.}
\begin{tabular}{|c|ccc|ccc|ccc|ccc|ccc|ccc|}
$W$(GeV)&\multicolumn{3}{c|}{$\phi=15^\circ$}
&\multicolumn{3}{c|}{$\phi=45^\circ$}
&\multicolumn{3}{c|}{$\phi=75^\circ$}
&\multicolumn{3}{c|}{$\phi=105^\circ$}
&\multicolumn{3}{c|}{$\phi=135^\circ$}
&\multicolumn{3}{c|}{$\phi=165^\circ$} \\ \hline
0.99& 16.6&$\pm$  10.1&$\pm$  6.3& 66&$\pm$  41&$\pm$  93&231&$\pm$  84&$\pm$ 101&200&$\pm$  70&$\pm$  77&243&$\pm$  59&$\pm$ 108&288&$\pm$  59&$\pm$ 104 \\
1.01& 14.9&$\pm$  5.4&$\pm$  5.6& 61&$\pm$  32&$\pm$  36&111&$\pm$  30&$\pm$  20&161&$\pm$  33&$\pm$  30&188&$\pm$  34&$\pm$  64&254&$\pm$  35&$\pm$  45 \\
1.03& 8.1&$\pm$  4.1&$\pm$  4.3& 45.2&$\pm$  14.9&$\pm$  6.4& 85&$\pm$  21&$\pm$  14&119&$\pm$  22&$\pm$  24&134&$\pm$  22&$\pm$  20&152&$\pm$  22&$\pm$  33 \\
1.05& 28.4&$\pm$  8.3&$\pm$  4.7& 41&$\pm$  13&$\pm$  15& 77&$\pm$  17&$\pm$  23&114&$\pm$  19&$\pm$  23&114&$\pm$  18&$\pm$  15& 83&$\pm$  15&$\pm$  62 \\
1.07& 14.6&$\pm$  6.8&$\pm$  4.0& 57.6&$\pm$  14.8&$\pm$  9.2& 59&$\pm$  13&$\pm$  13&104&$\pm$  17&$\pm$  18& 94&$\pm$  15&$\pm$  25&128&$\pm$  17&$\pm$  24 \\
1.09& 29.5&$\pm$  10.2&$\pm$  4.2& 39&$\pm$  10&$\pm$  10& 71.9&$\pm$  14.2&$\pm$  9.8& 68&$\pm$  12&$\pm$  10& 63&$\pm$  11&$\pm$  12& 67&$\pm$  12&$\pm$  24 \\
1.11& 14.4&$\pm$  6.4&$\pm$  5.0& 34.6&$\pm$  10.0&$\pm$  2.8& 66.9&$\pm$  12.8&$\pm$  9.2& 97&$\pm$  14&$\pm$  11&109.4&$\pm$  16.5&$\pm$  9.4& 71&$\pm$  13&$\pm$  18 \\
1.13& 36.7&$\pm$  11.2&$\pm$  7.7& 46.4&$\pm$  10.6&$\pm$  7.3& 59&$\pm$  11&$\pm$  15& 75.1&$\pm$  13.1&$\pm$  9.0& 82&$\pm$  15&$\pm$  17& 69.5&$\pm$  15.7&$\pm$  3.9 \\
1.15& 30.0&$\pm$  9.6&$\pm$  15.9& 82&$\pm$  14&$\pm$  15& 95.7&$\pm$  15.1&$\pm$  9.3& 79&$\pm$  14&$\pm$  14&125&$\pm$  20&$\pm$  43&105&$\pm$  18&$\pm$  27 \\
1.17& 54.1&$\pm$  12.9&$\pm$  9.5&103.1&$\pm$  16.1&$\pm$  8.9&104&$\pm$  17&$\pm$  12&124.4&$\pm$  17.5&$\pm$  9.3&180&$\pm$  22&$\pm$  24&168&$\pm$  22&$\pm$  10 \\
1.19& 94.3&$\pm$  16.4&$\pm$  7.5&118&$\pm$  17&$\pm$  13&135&$\pm$  17&$\pm$  13&154&$\pm$  18&$\pm$  23&145&$\pm$  19&$\pm$  14&178&$\pm$  24&$\pm$  15 \\
1.21&119&$\pm$  18&$\pm$  17&117&$\pm$  16&$\pm$  20&167.8&$\pm$  18.3&$\pm$  9.1&119&$\pm$  15&$\pm$  10&153.0&$\pm$  21.0&$\pm$  6.7&144&$\pm$  24&$\pm$  20 \\
1.23&111&$\pm$  16&$\pm$  13&108&$\pm$  15&$\pm$  18&102&$\pm$  13&$\pm$  11&141.6&$\pm$  16.8&$\pm$  8.0&107&$\pm$  17&$\pm$  18& 83&$\pm$  15&$\pm$  14 \\
1.25& 51&$\pm$  11&$\pm$  12& 78&$\pm$  12&$\pm$  11& 94&$\pm$  12&$\pm$  11& 74&$\pm$  13&$\pm$  10& 81.3&$\pm$  11.8&$\pm$  6.7& 96&$\pm$  12&$\pm$  10 \\
1.27& 41.3&$\pm$  9.2&$\pm$  7.1& 51.5&$\pm$  9.3&$\pm$  6.3& 48.1&$\pm$  9.8&$\pm$  6.9& 64.4&$\pm$  9.5&$\pm$  4.1& 61.4&$\pm$  8.3&$\pm$  4.8& 47.1&$\pm$  8.0&$\pm$  5.9 \\
1.29& 45.2&$\pm$  7.9&$\pm$  4.5& 40.3&$\pm$  7.6&$\pm$  7.1& 42.1&$\pm$  7.6&$\pm$  3.6& 40.1&$\pm$  5.5&$\pm$  3.9& 43.0&$\pm$  6.4&$\pm$  3.6& 35.0&$\pm$  8.2&$\pm$  5.0 \\
1.31& 29.7&$\pm$  7.0&$\pm$  15.2& 18.0&$\pm$  5.9&$\pm$  5.4& 42.3&$\pm$  6.0&$\pm$  2.4& 37.0&$\pm$  5.0&$\pm$  3.1& 41.4&$\pm$  7.7&$\pm$  3.5& 43.7&$\pm$  11.4&$\pm$  7.7 \\
1.33& 41.6&$\pm$  8.1&$\pm$  10.4& 33.0&$\pm$  6.1&$\pm$  3.1& 39.7&$\pm$  4.8&$\pm$  3.2& 31.1&$\pm$  5.1&$\pm$  3.4& 28.8&$\pm$  8.0&$\pm$  6.5& 20.9&$\pm$  8.6&$\pm$  1.4 \\
1.35& 19.9&$\pm$  4.8&$\pm$  4.4& 21.2&$\pm$  4.2&$\pm$  4.4& 30.2&$\pm$  4.0&$\pm$  2.1& 35.1&$\pm$  6.6&$\pm$  6.7& 27.1&$\pm$  8.3&$\pm$  8.0& 23.9&$\pm$  8.8&$\pm$  5.4 \\
1.37& 17.5&$\pm$  4.1&$\pm$  4.3& 21.6&$\pm$  3.7&$\pm$  2.1& 30.5&$\pm$  4.6&$\pm$  2.4& 11.8&$\pm$  5.0&$\pm$  9.5& 24.4&$\pm$  7.3&$\pm$  2.1& 22.5&$\pm$  9.6&$\pm$  8.2 \\
1.39& 18.0&$\pm$  3.8&$\pm$  3.1& 24.2&$\pm$  3.9&$\pm$  4.0& 18.5&$\pm$  4.7&$\pm$  4.4& 23.5&$\pm$  5.8&$\pm$  4.8& 29.1&$\pm$  8.5&$\pm$  6.8& 33.0&$\pm$  11.6&$\pm$  3.0 \\
1.41& 23.0&$\pm$  4.3&$\pm$  1.3& 20.3&$\pm$  4.2&$\pm$  2.0& 19.8&$\pm$  5.1&$\pm$  4.8& 31.8&$\pm$  6.2&$\pm$  5.1& 24.7&$\pm$  7.8&$\pm$  3.6& 32.7&$\pm$  10.0&$\pm$  4.4 \\
1.43& 23.9&$\pm$  4.8&$\pm$  2.3& 20.8&$\pm$  5.0&$\pm$  5.3& 16.8&$\pm$  4.5&$\pm$  3.0& 21.6&$\pm$  5.6&$\pm$  2.4& 33.7&$\pm$  8.0&$\pm$  4.7& 27.0&$\pm$  8.5&$\pm$  5.5 \\
1.45& 15.6&$\pm$  4.9&$\pm$  4.1& 24.8&$\pm$  6.1&$\pm$  3.2& 30.8&$\pm$  5.8&$\pm$  2.2& 33.4&$\pm$  6.6&$\pm$  4.5& 36.7&$\pm$  7.3&$\pm$  5.1& 45.5&$\pm$  10.4&$\pm$  6.7 \\
1.47& 34.3&$\pm$  7.4&$\pm$  2.9& 33.8&$\pm$  6.5&$\pm$  3.6& 37.0&$\pm$  6.2&$\pm$  2.3& 37.5&$\pm$  6.0&$\pm$  4.9& 44.8&$\pm$  8.0&$\pm$  6.0& 39.4&$\pm$  10.5&$\pm$  4.6 \\
1.49& 41.5&$\pm$  7.8&$\pm$  4.0& 41.5&$\pm$  6.8&$\pm$  4.2& 45.5&$\pm$  6.3&$\pm$  8.8& 45.4&$\pm$  5.9&$\pm$  7.0& 64.1&$\pm$  9.5&$\pm$  10.3& 67&$\pm$  11&$\pm$  10 \\
1.51& 36.5&$\pm$  6.7&$\pm$  4.6& 36.6&$\pm$  6.1&$\pm$  4.0& 48.6&$\pm$  5.6&$\pm$  5.9& 50.8&$\pm$  6.1&$\pm$  5.8& 47.6&$\pm$  7.4&$\pm$  5.2& 85&$\pm$  11&$\pm$  11 \\
1.53& 39.8&$\pm$  6.7&$\pm$  4.7& 39.9&$\pm$  6.0&$\pm$  5.4& 37.0&$\pm$  4.9&$\pm$  4.6& 42.4&$\pm$  5.6&$\pm$  4.5& 39.9&$\pm$  6.2&$\pm$  3.7& 50.4&$\pm$  8.7&$\pm$  6.3 \\
1.55& 37.1&$\pm$  6.3&$\pm$  2.5& 31.4&$\pm$  4.8&$\pm$  3.9& 26.8&$\pm$  4.0&$\pm$  4.6& 39.0&$\pm$  4.9&$\pm$  3.6& 57.8&$\pm$  7.0&$\pm$  5.2& 61.1&$\pm$  10.0&$\pm$  7.1 \\
1.57& 15.6&$\pm$  3.9&$\pm$  3.8& 19.7&$\pm$  3.8&$\pm$  3.5& 32.7&$\pm$  4.1&$\pm$  2.8& 36.8&$\pm$  4.3&$\pm$  3.2& 41.3&$\pm$  6.9&$\pm$  7.3& 31.7&$\pm$  9.2&$\pm$  10.2 \\
1.59& 18.5&$\pm$  3.9&$\pm$  1.3& 30.0&$\pm$  4.2&$\pm$  3.1& 31.5&$\pm$  3.8&$\pm$  3.5& 40.1&$\pm$  4.7&$\pm$  4.7& 27.8&$\pm$  6.4&$\pm$  5.0& 26.0&$\pm$  7.2&$\pm$  8.9 \\
1.61& 18.1&$\pm$  3.7&$\pm$  3.4& 21.7&$\pm$  3.7&$\pm$  3.5& 28.9&$\pm$  3.5&$\pm$  3.7& 23.8&$\pm$  4.5&$\pm$  3.5& 31.8&$\pm$  5.9&$\pm$  2.4& 26.2&$\pm$  6.4&$\pm$  4.0 \\
1.63& 19.0&$\pm$  4.0&$\pm$  2.2& 26.4&$\pm$  3.8&$\pm$  1.8& 29.4&$\pm$  3.8&$\pm$  2.6& 27.0&$\pm$  4.4&$\pm$  4.3& 32.6&$\pm$  5.2&$\pm$  4.2& 35.8&$\pm$  6.7&$\pm$  2.7 \\
1.65& 20.4&$\pm$  4.1&$\pm$  3.3& 25.1&$\pm$  3.7&$\pm$  2.2& 22.3&$\pm$  3.6&$\pm$  3.8& 29.2&$\pm$  3.8&$\pm$  2.4& 39.5&$\pm$  5.6&$\pm$  5.0& 36.8&$\pm$  7.5&$\pm$  6.8 \\
1.67& 17.7&$\pm$  3.6&$\pm$  3.4& 16.2&$\pm$  3.2&$\pm$  1.6& 26.1&$\pm$  3.5&$\pm$  2.2& 18.8&$\pm$  3.1&$\pm$  2.4& 17.0&$\pm$  5.3&$\pm$  4.7& 19.7&$\pm$  8.3&$\pm$  3.2 \\
1.69& 10.8&$\pm$  3.2&$\pm$  2.6& 16.2&$\pm$  3.4&$\pm$  3.6& 21.5&$\pm$  3.0&$\pm$  4.5& 16.0&$\pm$  3.6&$\pm$  2.9& 30&$\pm$  13&$\pm$  26& 22.4&$\pm$  14.6&$\pm$  9.8 \\
1.71& 13.1&$\pm$  3.7&$\pm$  3.8& 13.3&$\pm$  3.1&$\pm$  1.5& 14.1&$\pm$  2.6&$\pm$  3.5& 8.9&$\pm$  4.3&$\pm$  1.9& 16.8&$\pm$  5.6&$\pm$  2.3& 9.5&$\pm$  5.0&$\pm$  2.5 \\
1.73& 3.4&$\pm$  2.4&$\pm$  3.0& 6.2&$\pm$  2.4&$\pm$  4.5& 8.4&$\pm$  2.8&$\pm$  2.0& 14.4&$\pm$  3.6&$\pm$  2.0& 16.7&$\pm$  3.4&$\pm$  3.4& 14.3&$\pm$  3.7&$\pm$  2.3 \\
1.75& 17.2&$\pm$  3.5&$\pm$  2.0& 6.5&$\pm$  2.5&$\pm$  1.9& 11.1&$\pm$  3.3&$\pm$  1.3& 10.3&$\pm$  2.2&$\pm$  2.6& 6.7&$\pm$  2.4&$\pm$  1.3& 8.2&$\pm$  3.0&$\pm$  1.4 \\
1.77& 4.2&$\pm$  2.9&$\pm$  5.4& 5.1&$\pm$  2.5&$\pm$  0.9& 7.1&$\pm$  2.2&$\pm$  1.7& 8.0&$\pm$  1.7&$\pm$  1.0& 8.0&$\pm$  2.4&$\pm$  2.3& 11.6&$\pm$  3.4&$\pm$  1.6 \\
1.79& 5.6&$\pm$  2.6&$\pm$  3.3& 9.6&$\pm$  2.6&$\pm$  1.8& 7.0&$\pm$  1.6&$\pm$  1.8& 8.4&$\pm$  1.7&$\pm$  0.8& 3.7&$\pm$  4.0&$\pm$  3.5& 4.6&$\pm$  4.0&$\pm$  3.5 \\
1.81& 4.9&$\pm$  2.0&$\pm$  2.0& 8.8&$\pm$  2.0&$\pm$  0.9& 7.9&$\pm$  1.5&$\pm$  0.8& 13.6&$\pm$  2.2&$\pm$  3.1& 19.0&$\pm$  9.3&$\pm$  10.8& 18.6&$\pm$  9.3&$\pm$  9.8 \\
1.83& 8.2&$\pm$  1.9&$\pm$  1.1& 5.7&$\pm$  1.6&$\pm$  1.1& 7.7&$\pm$  1.6&$\pm$  0.9& 5.6&$\pm$  3.1&$\pm$  5.1& 4.9&$\pm$  3.6&$\pm$  2.1& 7.1&$\pm$  4.3&$\pm$  2.8 \\
1.85& 4.6&$\pm$  1.7&$\pm$  0.9& 8.4&$\pm$  1.8&$\pm$  1.5& 5.6&$\pm$  1.8&$\pm$  1.7& 7.3&$\pm$  2.4&$\pm$  1.1& 9.2&$\pm$  2.7&$\pm$  1.0& 8.3&$\pm$  3.2&$\pm$  2.2 \\
1.87& 4.6&$\pm$  1.9&$\pm$  1.2& 4.8&$\pm$  2.1&$\pm$  1.2& 5.4&$\pm$  2.0&$\pm$  1.6& 6.4&$\pm$  1.8&$\pm$  0.8& 7.1&$\pm$  2.3&$\pm$  2.3& 6.6&$\pm$  2.9&$\pm$  1.4 \\
1.89& 8.1&$\pm$  3.5&$\pm$  11.2& 8.7&$\pm$  2.6&$\pm$  2.2& 8.2&$\pm$  1.8&$\pm$  0.9& 5.8&$\pm$  1.6&$\pm$  1.0& 13.5&$\pm$  2.9&$\pm$  2.8& 7.8&$\pm$  3.6&$\pm$  1.9 \\
1.91& 10.1&$\pm$  2.6&$\pm$  1.9& 4.2&$\pm$  2.7&$\pm$  1.8& 5.0&$\pm$  1.4&$\pm$  1.2& 7.3&$\pm$  1.8&$\pm$  1.8& 5.7&$\pm$  3.7&$\pm$  1.8& 10.0&$\pm$  5.9&$\pm$  3.9 \\
\end{tabular}\end{ruledtabular}\end{table*}
\endgroup
\begingroup\squeezetable
\begin{table*}\begin{ruledtabular}
\caption{\label{table:ctcm2}$e p \rightarrow e p \gamma$ cross section ($\pm$ stat $\pm$ syst)
         at $Q^2$=1.0~GeV$^2$ and $\cos\theta_{\gamma\gamma}^\ast = -0.875$, in fm/MeV/sr$^2$.}
\begin{tabular}{|c|ccc|ccc|ccc|ccc|ccc|ccc|}
$W$(GeV)&\multicolumn{3}{c|}{$\phi=15^\circ$}
&\multicolumn{3}{c|}{$\phi=45^\circ$}
&\multicolumn{3}{c|}{$\phi=75^\circ$}
&\multicolumn{3}{c|}{$\phi=105^\circ$}
&\multicolumn{3}{c|}{$\phi=135^\circ$}
&\multicolumn{3}{c|}{$\phi=165^\circ$} \\ \hline
0.99&103&$\pm$  74&$\pm$  21&548&$\pm$ 306&$\pm$ 112&362&$\pm$ 159&$\pm$  75&459&$\pm$ 107&$\pm$  63&453&$\pm$  78&$\pm$  46&478&$\pm$  71&$\pm$  39 \\
1.01& 76&$\pm$  43&$\pm$  10&230&$\pm$  94&$\pm$  94&320&$\pm$  75&$\pm$  54&342&$\pm$  54&$\pm$  30&361&$\pm$  47&$\pm$  42&291&$\pm$  37&$\pm$  28 \\
1.03& 65&$\pm$  27&$\pm$  14&186&$\pm$  60&$\pm$  26&263&$\pm$  48&$\pm$  33&224&$\pm$  34&$\pm$  17&280&$\pm$  34&$\pm$  20&282&$\pm$  32&$\pm$  22 \\
1.05& 41&$\pm$  24&$\pm$  17&144&$\pm$  43&$\pm$  13&204&$\pm$  35&$\pm$  23&197&$\pm$  28&$\pm$  14&144&$\pm$  22&$\pm$  18&141&$\pm$  21&$\pm$  11 \\
1.07& 22.2&$\pm$  16.2&$\pm$  2.0&124&$\pm$  34&$\pm$  18& 87.7&$\pm$  19.6&$\pm$  9.0&154&$\pm$  22&$\pm$  12&134.5&$\pm$  19.4&$\pm$  6.5&157&$\pm$  20&$\pm$  14 \\
1.09& 16.9&$\pm$  12.1&$\pm$  4.5&115&$\pm$  29&$\pm$  10&143.9&$\pm$  23.4&$\pm$  9.5& 96&$\pm$  16&$\pm$  40&124&$\pm$  17&$\pm$  14&121&$\pm$  19&$\pm$  11 \\
1.11& 30.7&$\pm$  15.7&$\pm$  7.7& 69.9&$\pm$  20.7&$\pm$  4.2&106.7&$\pm$  18.4&$\pm$  8.3&134&$\pm$  18&$\pm$  10&129&$\pm$  21&$\pm$  13& 81&$\pm$  20&$\pm$  45 \\
1.13& 30.1&$\pm$  14.1&$\pm$  5.2& 85.5&$\pm$  20.1&$\pm$  8.1& 82.8&$\pm$  14.8&$\pm$  6.6&114.0&$\pm$  17.9&$\pm$  4.5& 78&$\pm$  19&$\pm$  16&105&$\pm$  30&$\pm$  11 \\
1.15& 59.7&$\pm$  20.1&$\pm$  3.9& 45&$\pm$  13&$\pm$  12&133.1&$\pm$  19.0&$\pm$  9.7& 99&$\pm$  18&$\pm$  18& 98&$\pm$  22&$\pm$  13&116&$\pm$  25&$\pm$  10 \\
1.17& 31.1&$\pm$  14.1&$\pm$  8.2& 43&$\pm$  14&$\pm$  15&110.2&$\pm$  17.7&$\pm$  5.1&107&$\pm$  18&$\pm$  11&118&$\pm$  22&$\pm$  14&106&$\pm$  30&$\pm$  14 \\
1.19&130&$\pm$  28&$\pm$  26&118.5&$\pm$  18.9&$\pm$  8.4&123&$\pm$  17&$\pm$  11&126&$\pm$  18&$\pm$  12&145&$\pm$  29&$\pm$  18&272&$\pm$ 122&$\pm$  90 \\
1.21&103&$\pm$  21&$\pm$  11&124.6&$\pm$  19.0&$\pm$  7.2&135.7&$\pm$  17.5&$\pm$  5.5&183&$\pm$  21&$\pm$  11&177&$\pm$  51&$\pm$  18&217&$\pm$  70&$\pm$  46 \\
1.23&129.0&$\pm$  21.1&$\pm$  8.7&117.4&$\pm$  18.5&$\pm$  6.7&135.2&$\pm$  16.3&$\pm$  9.8&133&$\pm$  20&$\pm$  12&130&$\pm$  24&$\pm$  42& 68&$\pm$  19&$\pm$  28 \\
1.25& 99.1&$\pm$  17.4&$\pm$  6.2& 95.9&$\pm$  17.8&$\pm$  7.0& 99.2&$\pm$  13.4&$\pm$  7.4& 68.2&$\pm$  14.1&$\pm$  8.3& 88&$\pm$  14&$\pm$  14& 81.8&$\pm$  21.3&$\pm$  8.8 \\
1.27& 54.1&$\pm$  13.9&$\pm$  7.6& 60&$\pm$  13&$\pm$  14& 87.1&$\pm$  13.2&$\pm$  5.6& 71.3&$\pm$  11.8&$\pm$  6.3& 58.0&$\pm$  12.5&$\pm$  7.3&&& \\
1.29& 35.6&$\pm$  11.8&$\pm$  5.7& 41.8&$\pm$  10.1&$\pm$  9.8& 47.8&$\pm$  10.0&$\pm$  4.1& 70.0&$\pm$  9.4&$\pm$  4.7& 58.5&$\pm$  19.2&$\pm$  7.6&&& \\
1.31& 29.6&$\pm$  10.7&$\pm$  5.7& 51.8&$\pm$  10.7&$\pm$  5.9& 49.2&$\pm$  9.0&$\pm$  3.1& 58.3&$\pm$  8.8&$\pm$  3.7&&&&&& \\
1.33& 29.3&$\pm$  8.5&$\pm$  2.5& 44.8&$\pm$  10.4&$\pm$  4.7& 35.0&$\pm$  6.6&$\pm$  4.3& 39.8&$\pm$  8.8&$\pm$  2.7&&&&&& \\
1.35& 10.4&$\pm$  5.0&$\pm$  7.7& 28.9&$\pm$  9.7&$\pm$  4.6& 31.2&$\pm$  5.4&$\pm$  3.3& 42&$\pm$  11&$\pm$  10&&&&&& \\
1.37& 24.0&$\pm$  9.9&$\pm$  5.9& 57.9&$\pm$  12.3&$\pm$  5.0& 45.0&$\pm$  6.3&$\pm$  4.1& 38.4&$\pm$  12.4&$\pm$  4.1&&&&&& \\
1.39&&&& 42.1&$\pm$  7.5&$\pm$  7.4& 35.9&$\pm$  6.2&$\pm$  2.3& 31.8&$\pm$  10.3&$\pm$  6.8&&&&&& \\
1.41&&&& 21.3&$\pm$  4.8&$\pm$  7.8& 30.8&$\pm$  6.9&$\pm$  1.8& 18.0&$\pm$  9.1&$\pm$  4.1&&&&&& \\
1.43&&&& 32.1&$\pm$  5.7&$\pm$  4.2& 27.6&$\pm$  8.0&$\pm$  6.6& 25.8&$\pm$  11.0&$\pm$  7.0&&&&&& \\
1.45&&&& 36.9&$\pm$  7.3&$\pm$  8.8& 31.3&$\pm$  8.6&$\pm$  5.5& 11.4&$\pm$  8.4&$\pm$  10.1&&&&&& \\
1.47&&&& 44.4&$\pm$  11.7&$\pm$  8.6& 25.3&$\pm$  7.9&$\pm$  10.7& 49.3&$\pm$  12.4&$\pm$  9.3&&&&&& \\
1.49&&&& 47&$\pm$  13&$\pm$  14& 46.8&$\pm$  10.0&$\pm$  6.9& 70.4&$\pm$  14.3&$\pm$  8.9&&&&&& \\
1.51&&&& 38.4&$\pm$  10.8&$\pm$  6.5& 57.1&$\pm$  10.6&$\pm$  5.7& 50&$\pm$  12&$\pm$  16&&&&&& \\
1.53&&&& 40.5&$\pm$  11.4&$\pm$  6.8& 53.4&$\pm$  10.0&$\pm$  3.4& 37&$\pm$  10&$\pm$  12&&&&&& \\
1.55&&&& 30&$\pm$  13&$\pm$  16& 45.2&$\pm$  8.4&$\pm$  2.6& 42.0&$\pm$  10.9&$\pm$  6.5&&&&&& \\
1.57&&&& 10.3&$\pm$  8.8&$\pm$  1.8& 36.7&$\pm$  7.8&$\pm$  6.8& 59.3&$\pm$  12.3&$\pm$  5.2&&&&&& \\
1.59&&&& 15.5&$\pm$  7.3&$\pm$  9.9& 22.1&$\pm$  6.7&$\pm$  3.6& 37.2&$\pm$  12.6&$\pm$  4.3&&&&&& \\
1.61&&&& 12.8&$\pm$  6.7&$\pm$  6.5& 28.4&$\pm$  7.4&$\pm$  3.3& 30.8&$\pm$  11.8&$\pm$  3.7&&&&&& \\
1.63&&&&&&& 47.3&$\pm$  9.1&$\pm$  4.3& 32.9&$\pm$  12.4&$\pm$  6.7&&&&&& \\
1.65&&&&&&& 48.9&$\pm$  9.7&$\pm$  6.0& 27.3&$\pm$  14.4&$\pm$  5.3&&&&&& \\
1.67&&&&&&& 18.1&$\pm$  7.4&$\pm$  7.1& 47&$\pm$  13&$\pm$  15&&&&&& \\
\end{tabular}\end{ruledtabular}\end{table*}
\endgroup
\begingroup\squeezetable
\begin{table*}\begin{ruledtabular}
\caption{\label{table:ctcm3}$e p \rightarrow e p \gamma$ cross section ($\pm$ stat $\pm$ syst)
         at $Q^2$=1.0~GeV$^2$ and $\cos\theta_{\gamma\gamma}^\ast = -0.650$, in fm/MeV/sr$^2$.}
\begin{tabular}{|c|ccc|ccc|ccc|ccc|ccc|ccc|}
$W$(GeV)&\multicolumn{3}{c|}{$\phi=15^\circ$}
&\multicolumn{3}{c|}{$\phi=45^\circ$}
&\multicolumn{3}{c|}{$\phi=75^\circ$}
&\multicolumn{3}{c|}{$\phi=105^\circ$}
&\multicolumn{3}{c|}{$\phi=135^\circ$}
&\multicolumn{3}{c|}{$\phi=165^\circ$} \\ \hline
0.99&&&&&&&554&$\pm$ 216&$\pm$  88&452&$\pm$  90&$\pm$  38&454&$\pm$  64&$\pm$  23&340&$\pm$  49&$\pm$  26 \\
1.01&&&&&&&380&$\pm$  87&$\pm$  71&286&$\pm$  46&$\pm$  17&213&$\pm$  30&$\pm$  16&274&$\pm$  31&$\pm$  17 \\
1.03&&&&237&$\pm$ 109&$\pm$  55&270&$\pm$  50&$\pm$  18&195&$\pm$  28&$\pm$  16&233&$\pm$  27&$\pm$  12&205&$\pm$  25&$\pm$  11 \\
1.05&&&&277&$\pm$  94&$\pm$  63&234&$\pm$  41&$\pm$  12&211&$\pm$  27&$\pm$  16&167&$\pm$  22&$\pm$  14&173.9&$\pm$  22.2&$\pm$  9.6 \\
1.07&&&& 65&$\pm$  36&$\pm$  19&183.2&$\pm$  29.7&$\pm$  8.7&161&$\pm$  21&$\pm$  10&140&$\pm$  18&$\pm$  11&142&$\pm$  19&$\pm$  20 \\
1.09&&&&266&$\pm$  59&$\pm$  12&153.9&$\pm$  24.0&$\pm$  9.5&169.1&$\pm$  20.6&$\pm$  9.2&138&$\pm$  19&$\pm$  10& 97&$\pm$  21&$\pm$ 101 \\
1.11&&&&189&$\pm$  44&$\pm$  18&139.0&$\pm$  21.1&$\pm$  5.3&108.1&$\pm$  15.8&$\pm$  7.7&152&$\pm$  26&$\pm$  41&&& \\
1.13&462&$\pm$ 181&$\pm$ 270&195.4&$\pm$  41.0&$\pm$  9.0&113.5&$\pm$  17.1&$\pm$  5.6&123.1&$\pm$  18.8&$\pm$  7.5&&&&&& \\
1.15&131&$\pm$  69&$\pm$  97& 92.1&$\pm$  26.4&$\pm$  9.9&102.4&$\pm$  15.6&$\pm$  5.4&125&$\pm$  20&$\pm$  11&&&&&& \\
1.17&&&&132&$\pm$  27&$\pm$  13& 95&$\pm$  15&$\pm$  10&135&$\pm$  21&$\pm$  15&&&&&& \\
1.19&&&&107&$\pm$  22&$\pm$  24&168.7&$\pm$  20.3&$\pm$  5.9&152&$\pm$  22&$\pm$  27&&&&&& \\
1.21&&&&154&$\pm$  24&$\pm$  22&153&$\pm$  18&$\pm$  14&125&$\pm$  21&$\pm$  17&&&&&& \\
1.23&&&&104&$\pm$  19&$\pm$  16&132&$\pm$  17&$\pm$  10& 87&$\pm$  21&$\pm$  10&&&&&& \\
1.25&&&&108&$\pm$  19&$\pm$  12& 76&$\pm$  13&$\pm$  12& 50&$\pm$  15&$\pm$  59&&&&&& \\
1.27&&&&107&$\pm$  20&$\pm$  11&103&$\pm$  15&$\pm$  15& 85&$\pm$  16&$\pm$  73&&&&&& \\
1.29&&&& 51.1&$\pm$  17.4&$\pm$  6.8& 79&$\pm$  15&$\pm$  15& 43&$\pm$  10&$\pm$  12&&&&&& \\
1.31&&&& 77&$\pm$  19&$\pm$  23& 69&$\pm$  13&$\pm$  14& 68&$\pm$  14&$\pm$  32&&&&&& \\
1.33&&&& 60&$\pm$  16&$\pm$  16& 42&$\pm$  10&$\pm$  15& 54&$\pm$  15&$\pm$  42&&&&&& \\
1.35&&&& 62.7&$\pm$  16.3&$\pm$  4.4& 65.6&$\pm$  11.4&$\pm$  3.4&&&&&&&&& \\
\end{tabular}\end{ruledtabular}\end{table*}
\endgroup
%
\bibliography{common}
\end{document}